\documentclass{article}

\usepackage{tenor}
\usepackage{ifpdf}
\usepackage[english]{babel}
\usepackage{stmaryrd} 

\let\xcup\cup
\usepackage{version}

\def\papertitle{
  Härpfer's Extended Indispensability Algorithm in Z
}
\def\firstauthor{Markus Lepper}
\def\secondauthor{Bernd Härpfer}
\def\thirdauthor{Baltasar Trancón y Widemann}

\def\copyrightyear{2024}

\newif\ifpdf
\ifx\pdfoutput\relax
\else
   \ifcase\pdfoutput
      \pdffalse
   \else
      \pdftrue
\fi

\ifpdf 
  \usepackage[pdftex,
    pdftitle={\papertitle},
    pdfauthor={\firstauthor},
    bookmarksnumbered, 
    pdfstartview=XYZ 
   ]{hyperref}
\fi
\usepackage[pdftex]{graphicx}

\usepackage{amsmath}
\usepackage{amssymb}
\usepackage{amsthm}

  \usepackage[figure,table]{hypcap}

\fi

\hypersetup{
    colorlinks,%
    citecolor=black,%
    filecolor=black,%
    linkcolor=black,%
    urlcolor=black
}

\title{\papertitle}
 \threeauthors
   {\firstauthor}{semantics gGmbH, Berlin}
   {\secondauthor}{\url{haerpfer.net}}
   {\thirdauthor}{Technische Hochschule Brandenburg\\semantics gGmbH, Berlin}

\newcommand{\xumbi}[1]{#1}

\makeatletter
\thm@headpunct{X}
\newtoks\thm@headpunct \thm@headpunct{X}
\makeatother

\renewcommand\emptyset{\varnothing}

\newcommand{\makeobjectTT}[1]%
{\expandafter\newcommand\csname x#1\endcsname{\ensuremath{\mathord{\mathtt{#1}}}}}
\newcommand{\makeobjectIT}[1]%
{\expandafter\newcommand\csname x#1\endcsname{\ensuremath{\mathord{\mathit{#1}}}}}
\newcommand{\makeobjectSF}[1]%
{\expandafter\newcommand\csname x#1\endcsname{\ensuremath{\mathord{\mathsf{#1}}}}}
\newcommand{\makeobjectBF}[1]%
{\expandafter\newcommand\csname x#1\endcsname{\ensuremath{\mathord{\mathbf{#1}}}}}

\newcommand{\makeobjectITop}[1]%
{\expandafter\newcommand\csname x#1\endcsname{\ensuremath{\operatorname{\mathit{#1}}}}}
\newcommand{\makeobjectSFop}[1]%
{\expandafter\newcommand\csname x#1\endcsname{\ensuremath{\operatorname{\mathsf{#1}}}}}

\makeobjectBF{ID}
\makeobjectBF{if}
\makeobjectBF{then}
\makeobjectBF{else}

\newcommand{\xfrac}[2]{\frac{\begin{array}{@{}c}#1\end{array}}%
{\begin{array}{@{}c}#2\end{array}}}
\newcommand{\nat}{\ensuremath{\mathbb{N}}} 

\newcommand{\limg}{\ensuremath{\mathopen{(\kern-0.4ex|}}}
\newcommand{\rimg}{\ensuremath{\mathclose{|\kern-0.4ex)}}}

\newcommand{\power}{\ensuremath{\operatorname{\mathbb{P}}}}

\newcommand{\pfun}{\ensuremath{\mathrel\nrightarrow}}
\newcommand{\fun}{\ensuremath{\mathrel\rightarrow}}

\newcommand{\dom}{\ensuremath{\mathop{{}\mathsf{dom}}}}
\newcommand{\ran}{\ensuremath{\mathop{{}\mathsf{ran}}}}
\newcommand{\compose}{\ensuremath{\mathbin{\fatsemi}}}
\newcommand{\chop}{\ensuremath{\smallfrown}}

\newcommand{\rinv}{\ensuremath{^\sim}} 
\newcommand{\domres}{\ensuremath{\triangleleft}}
\newcommand{\ranres}{\ensuremath{\triangleright}}

\makeobjectSFop{min}
\makeobjectSFop{max}
\makeobjectSF{mod}

\makeobjectSFop{seq} 
\makeobjectITop{head}
\makeobjectITop{tail}

\newcommand{\xcomm}[1]{{\small\textit{// #1}}}

\makeobjectSF{GNSM}
\makeobjectSF{gnsm}
\makeobjectSF{MNSM}
\makeobjectSF{NSM}
\makeobjectSF{getIndispensability}
\makeobjectSF{descend}
\makeobjectSF{occurences}
\makeobjectSF{topPulses}
\makeobjectSF{strata}
\makeobjectSF{topValues}
\makeobjectSF{startPositions}
\makeobjectIT{sort}
\makeobjectIT{cyclicSucc}
\makeobjectIT{squash}
\makeobjectSF{combine}

\makeobjectSF{expand}
\makeobjectSF{insert}
\makeobjectSF{append}
\makeobjectSF{fillLevel}
\newcommand{\dotdot}{\ensuremath\mathrel{..}}

\begin{document}
%
\capstartfalse
\maketitle
\capstarttrue

\section{Context}

Since 1978, Clarence Barlow developed the ``Indispensability Function'' (see
\cite{barlow80}, pages 35 ff. and 82, \cite{barlow87}, pages 56--57).
A very simplified explanation is as follows:
The function operates on a metric tree that is bound to the same prime number
of branches for all subtrees of each particular level.
It assigns to all leaf postions of this tree
a numeric value which indicates how important the acoustic
presence of an event at this position is for the meter to be
recognized as such.
For each position a different integer value is calculated
and the values range from $0$ to the number of positions minus $1$
(For a thorough discussion see \cite{barlow12, haerpfer}).

Bernd Härpfer extended this concept in 2015, and wrote a new algorithm
to deal with meters (metric trees) which have arbitrary groupings
into two or three (of single positions or of arbitrary complex sub-trees)
at any position of the tree hierarchy, now called
``Extended Indispensability Algorithm''.

Härpfer gives (a) a description of his algorithm in human language,
providing examples in section 4.3.3 of his thesis \cite{haerpfer},
and gives (b) an implementation in C++, in the appendix of that
work.

The following text gives (c) a specification in (a slightly extended
version) of the Z specification language. \cite{spivey-z} Once the
reader is accustomed to the basic semantic concepts and their
notation, this is a lean, unambiguous, and mathematically
well-founded way of specification. Its main advantages in reading
and understanding are that the semantics are completely functional:
The values of variables are constructed step by step and are never
overwritten. The toolkit for manipulating relations allows to
notate in short expressions what needs several lines of code in
imperative programming languages. Nevertheless, specifying in Z is
basically also a smart way of programming.

\section{Explanation of the Algorithm and the Formalism}

\subsection{Generic notation for stratified meters}

\begin{table}[t]
\centering
  $\begin{array}{c}

  \_\circ\_ :  \power\xseq\nat \times
   \power\xseq\nat \fun \power\xseq\nat \\
  a \circ b = \{ v \in a, w \in b \bullet v \chop w \}\\

  \_\uparrow\_ :  \power\xseq\nat \times \nat \fun
              \power\xseq\nat \\
  a \uparrow 0  =  \{\langle\rangle\} \\
  n > 0 \Longrightarrow 
  a \uparrow n  =  a \circ (a \uparrow  (n-1)) \\[2ex]

  \xGNSM, \xMNSM : \nat \times \nat \times \nat 
  \fun \power\xseq\nat\\
  \xgnsm  : (\nat \cup\{-1\}) \times \nat \times \nat 
  \fun \power\xseq\nat\\[1.5ex]

  \xgnsm_{-1,\_,\_} =  \{\langle \rangle\} \\[2ex]

  \xfrac{f \geq 0
    }{
 \xgnsm_{f,i,a} = \xGNSM_{f,i,a} \\
 \xGNSM_{f,i,a} = \bigcup   n : \nat  \mid  i \leq n \leq a 
 \bullet  \xMNSM_{f,i,a} \uparrow n  \\
 \xMNSM_{f,i,a} =  \{ u \in \xgnsm_{f-1,i,a}  \bullet u \oplus \langle f \rangle\}
 \\[3ex]
 \xGNSM = \bigcup f, i, a : \nat  \bullet \xGNSM_{f,i,a}  \\
  \xMNSM = \bigcup f, i, a : \nat  \bullet \xMNSM_{f,i,a}  \\
  \xNSM = \xGNSM \cup \xMNSM  \\
}  

\end{array}$
\caption{Bottom-up construction of the generic and the measure
  notation for stratified meters
  (GNSM and MSNM)
  \label{tab-gnsm-bottom-up}}
\end{table}

The input data to Härpfer's algorithm 
are sequences of integer values which make up a
\emph{generic notation for stratified meters (GNSM)}.  This is a compact
notation for a metric tree and specifies for each position in the meter the
level, on which this time point is the root of a subtree, see section~4.2.3
in \cite{haerpfer}.

Each GNSM is a sequence of integers from a set $\{0\dotdot f\}$.
These stand for \emph{metric weights}. 
The value $f$ stands for the convergence points of the highest
metric level; decreasing numbers stand for the lower metric levels,
down to $0$. The parameters $i$ and $a$ give the minimal and the
maximal number of positions on the top-level of each
sub-tree.

A bottom-up construction for GNSM is found in 
Table~\ref{tab-gnsm-bottom-up}:
The first lines define two auxiliary functions not provided by the
standard Z toolkit: $\_\circ\_$ combines two sets of sequences
(= ``languages'' in computer science speak) by concatenating all
possible combinations; $\_\uparrow\_$ multiplies these sequences by
multiple concatenations.

The case $f = -1$ is not a sensible use case but required
as the base case of construction: $\xgnsm_{-1,\_, \_}$
produces a set containing only an empty sequence.
Each set of sequences $\xMNSM_{s,i,a}$ represents possible \emph{subtrees},
in which the very first position is increased to the level $s$,
to mark the start of a metric substructure.

(In this paper sequences of length $k$ are realized by functions with a
contiguous domain $\{0\dotdot k-1\}$.
The sequences in the standard Z toolkit are 1-based.
The operator $\oplus$ means point-wise
overriding one relation by another.)

Each $\xGNSM_{s,i,a}$ contains all repetitions of these sub-trees,
with the number of repetitions in the given limits $i$ and $a$.

For instance: $\xGNSM_{-1, 2, 3} = \{\langle\rangle\}$, and thus 
$\xMNSM_{0, 2, 3} = \{\langle0\rangle\}$.
Then  $\xGNSM_{0, 2, 3}$ is the set of all allowed repetitions
$ = \{\langle0, 0\rangle, \langle0, 0, 0\rangle\}$
and $\xMNSM_{1, 2, 3} = \{\langle1, 0\rangle, \langle1, 0, 0\rangle\}$
are those very sequences marked as subtrees.

Finally   $\xGNSM_{1, 2, 3} $\xumbi{\\}$ = \{\langle1, 0, 1, 0\rangle,
\langle1, 0, 0, 1, 0 \rangle
\langle1, 0, 1, 0, 0\rangle
\langle1, 0, 0, 1, 0, 0\rangle
\}$ \xumbi{\\} contains all possible combinations of these subtrees.

Depending on the further processing, the values for $i$ and
$a$ can be chosen differently.
The input to Härpfer's algorithm are data from
$\xGNSM_{\_,2,3}$: every point is followed by one or two points
of the next lower metric weight, before the same or a higher
metric weight appears.
This is proposed and discussed in detail on pages~150 ff. of
\cite{haerpfer} in its historic context with reference (among others)
to Lehrdahl and Jackendoff's metric well-formedness rule ``MWFR 3''
\cite{lerdahl1983a-generative-th}.

Other data types may also be sensible:
$\xGNSM_{\_,1,3}$ allows two positions of the same metric weight to be adjacent
on a level higher than zero;
$\xGNSM_{\_,2,5}$ allows five pulses of equal weight, etc.

\subsection{MNSM, a Variant of GNSM}

Some irregularity in the specification for GNSM rises from the
fact that the top metric structure is represented by multiple
equal numbers. This is appropriate for the research carried out in
\cite{haerpfer}, but not always.
When the metric structure is intended to be executed identically
repeated, namely for a sequence of \emph{measures}, and when the start
of each measure can be received as such by the listener, it can be more
appropriate to mark this outstanding timepoint with a singular
metric weight. 

This is called \emph{measure notation for stratified meters}
\xumbi{\\}
 (MNSM) and is realized by the definition of $\xMNSM_{s,i,a}$ in
Table~\ref{tab-gnsm-bottom-up}.

The difference between both forms is to represent for instance
a conventional 6/8 meter as
\[ \langle 1 , 0 , 0 , 1 , 0 , 0 \rangle ~~\in~~\xGNSM_{1,2,3} \]\\[-6ex]
or as 
\[ \langle 2 , 0 , 0 , 1 , 0 , 0 \rangle ~~\in~~\xMNSM_{2,2,3} \]

\begin{table}[t]
\centering
  $\begin{array}{l}
\xsort : \power \nat \fun \xseq \nat \\
\xsort (A) = \xsquash(\xID_A) \\[1ex]

\xcyclicSucc : \power \nat \fun (\nat \pfun \nat) \\[1ex]
\xfrac{B = \xsort A
  }{
    \xcyclicSucc (A) = B\rinv\compose(\xtail B \chop \langle \xhead B\rangle)
  }\\[2ex]

  \xcombine : (\nat\pfun\nat) \times (\nat\pfun\nat)\fun (\nat\pfun\nat)
  \\
  \xcombine (A, B) = (A \compose (\_ + \#B)) \xcup B    \\[1ex]

  \xoccurences : \xNSM \times \nat \fun\power\nat \\
  \xoccurences(g, n) = g\rinv\limg\{n \}\rimg \\[2ex]

  \xgetIndispensability :   \xGNSM_{\_,2,3} \fun \xseq \nat \\

  \xdescend :   \xGNSM_{\_,2,3}  \times (\nat\pfun\nat)
    \fun (\nat \pfun \nat) \\[2ex]
    
    \xfrac{
      \xstrata = g(1) \\
      \xtopPulses  = \xoccurences(g, \xstrata) \\
      b = \xsort(\xtopPulses)\rinv \compose
      ~(\xif \#\xtopPulses = 2~\xthen~\langle 1, 0\rangle
 \\\hspace*{10em}           ~\xelse~\langle 2, 0, 1\rangle)
    }{
      \xgetIndispensability(g)  = \xdescend (g, b)\\
    }\\[4ex]
    \xfrac{
      f = \xmin  ( g \limg \dom c \rimg) - 1  \\
      F = \xoccurences(g, f) \\ 
      G = F \xcup \dom c \\
      H = F \domres \xcyclicSucc(G) \\
      \ran(H \compose H) \subset \dom c \\
      J = H \compose c \\
      K = H \compose J \\
      L = ( \xsquash (K{} \rinv) )\rinv\\
      c' = \xcombine(\xcombine(c, J), L) \\
    }{
      \xdescend(g, c) = \xif~ f=0 ~\xthen~ c' ~\xelse~\xdescend(g, c')
      } \\
  \end{array}$ 
\caption{Härpfer's Extended Indispensability Algorithm (\cite{haerpfer},
    Section 4.3.3, pages 174--177)}
  \label{tab-algo3}
\end{table}

\subsection{Extended Indispensability Algorithm}

The formulas in Table~\ref{tab-algo3} give a Z version of Härpfer's algorithm
for ``extended indispensability''.

The main function \xgetIndispensability{} calculates all indispensability
values.

The indexes in both input and output sequences stand for the same sequence
of time points or metric positions.
The input sequence is from $\xGNSM_{s,2,3}$ with $s\geq 0$,
as defined above,
and its values stand for the metric weights at these positions.
The values in the output
stand for the calculated indispensability values of the corresponding
positions.

The first equation of function \xgetIndispensability{} extracts the first
metric weight into the value \xstrata.
Due to the definition of \xGNSM, this is also the largest appearing value.

\xtopPulses{} is the set of all indexes where that highest weight
appears.
(Technically:
The function \xoccurences{} takes a \xGNSM{} or \xMNSM{}
input sequence as a relation, builds its inverse, and
applies this to the set containing only one particular weight.)

The list $b$ is the initial value of the constructed result list, as a
(partial) mapping from indexes (=~time points) to weights. Initial
indispensability values
are assigned explcitly (``hard-coded'') to the indexes in \xtopPulses{},
in two different ways, whether there are two or three of them.

The function \xdescend{} is an auxiliary function, operating stepwise and
recursively on the lower values of metric weights. Its additional parameter
$c$ contains the indispensability values of the higher metric weights, as
constructed so far.

The function first determines the level $f$ of the metric tree (given as
GNSM) which must be processed next.  ($f +1$ is the minimal value in the
GNSM for which an indispensability value has already been calculated.
Technically:
Take the result so far, as given by $c$,
  take all positions which it defines, take all
the metric weights at these positions, and take the minimum of
these.  The C-code calls this value the ``level in \emph{f}ocus'',
therefore we have chosen the abbreviation $f$).

The definition of \xdescend{} ensures for the metric weight in focus that
$f \geq 0$.  The set $F$ contains all indexes which carry that focused
metric weight (using the same technique as for \xtopPulses{} above.)

$G$ is the set of all time points which have the focused metric weight or a
higher one.  Because we descend through the metric weights, all time points
with a higher weight have already an indispensability value by the input
data $c$.

The time points in $F$ will be assigned an indispensability value in this
call to \xdescend{}.

$\xcyclicSucc$ is an auxiliary definition which maps
every element in a set of (natural) numbers to the next higher
number contained, and the highest number to the lowest.
(Technically: construct the sorted sequence, then invert it to
map each number to its position, then look up this position in
the modified list with the first element rotated to the end.)

The next line states that every successor of a successor has a higher
weight, which means that at most two values of the same weight may be
adjacent in $G$.  (This property is not stated by the original text
and follows from $i=2$ of the input data type.)

$J$ maps all the time points from $F$ which have a higher-level time point
as their successor to the value already assigned to this successor by
$c$.
(Technically: Those time points from $F$ which do not map to $\dom c$ but
to $F$ are ignored by the composition operator $\compose$.)

$\xcombine$ is an auxiliary function which takes two assignments of
positions to values. The positions in the first set shall have
the higher values. Therefore these are incremented by the size of the second
set. If the values in both assigments are compact sets
of the form $\{0\dotdot\#A -1\}$ and $\{0\dotdot\#B -1\}$, then the result
is also of this form.

This function is used to combine the sets $c$ and $J$ at the
end of Table~\ref{tab-algo3}.
The relation $H$ is an injective function and its range is a superset of
the domain of $c$.
(Because from $i=2$ it follows that every position from $c$ has at least
one (cyclic) predecessor from $F$.)
If the incoming value $c$ is compact then $J$ is compact, too.

Still unassigned are all time points with the focused weight $f$ which have
another such time point as its successor (= a successor from $F$ and not
from $\dom c$.  Each of these is a ``second period of a three period
group'' in the wording of the original description.)

$K$ maps these indexes to the indispensability values of their successors.
In contrast to $J$ its range is not necessarily compact,
caused by the fact that $i=2$ and $a=3$ allows sub-trees of different
breadth.
$L$ compactifies these values into the range $0\dotdot \# K -1$.  (Technically
done by converting the inverted relation to a sequence by 
our 0-based version of the ``Z toolkit
function'' \xsquash, and inverting again.)

\newcommand{\xp}{\swarrow}

This is the part most difficult to read in both the C++ and Z sources.
An example may help:

\[\begin{array}{l|*8c|c}
\mbox{position:}       & 1 & 2 & 3 & 4 & 5 & 6 & 7 & 8 & 1 \\
\hline
\mbox{GNSM} \hfill     g=  & 1 & 0 & 0 & 1 & 0 & 1 & 0 & 0 & 1 \\
\hline
\mbox{initially}\hfill c=
                       & 2 &   &   & 0 &   & 1 &   &   & 2 \\
\hline
                   &   &   &\xp&   &\xp&   &   &\xp& \\
\mbox{copy}\hfill  J=  &   &   & 0 &   & 1 &   &   & 2 & \\
\hline
                       &   &\xp&   &   &   &   &\xp&   & \\
\mbox{copy} \hfill K=  &   & 0 &   &   &   &   & 2 &   & \\
\mbox{compact}\hfill L=&   & 0 &   &   &   &   & 1 &   & \\
\hline
\xcombine(c,J) =       & 5 &   & 0 & 3 & 1 & 4 &   & 2 & \\
\mbox{result}\hfill c'=& 7 & 0 & 2 & 5 & 3 & 6 & 1 & 4 & \\
\end{array}\]
From $K$ to $L$ the relation $\{(2,0),(7,2)\}$ is inverted to 
$\{(0,2),(2,7)\}$, squashed to $\{(0,2),(1,7)\}$, and again inverted
to $\{(2,0),(7,1)\}$.

If we have not yet reached metric weight $0$, the function
$\xdescend{}(..)$  is
called recursively. Otherwise the assignments collected so far are returned
as a result.

Formally the accumulator parameter $c$
is only defined as a
partial function, i.e.\ a special subtype of relation. That it is
finally an injective sequence, i.e.\ a function with the domain 
$\{0\dotdot \#g-1\}$ is only stated implicitly by the function
signature of \xgetIndispensability.

This, and the fact that also its range has this value, follows from the
properties of \xdescend: the result of every application assigns
to all positions from $F$ completely, the values taken by $f$
during the recursive applications cover all
values in the range of $g$, and all results $c'$, the intermediate and
the final, are compact with $\ran c' = \{0\dotdot \# c' -1\}$.

\begin{table}[t]
\centering
  $\begin{array}{l}

  \xgetIndispensability :
  \xNSM  \fun \xseq \nat \\
  \xstartPositions : \nat \pfun \xseq \nat\\
    \xdescend :
  \xNSM  \times \xseq\nat \fun \xseq\nat  \\[2ex]

\xstartPositions(1) = \langle 0\rangle\\
\xstartPositions(2) = \langle 0, 1\rangle\\
\xstartPositions(3) = \langle 0, 2, 1 \rangle\\
\xstartPositions(4) = \langle 0, 2, 1, 3 \rangle
   ~\xcomm{still questionable, see text. }\\
\xcomm{~~etc.}\\[2ex]

    \xfrac{
      \xstrata = g(1) \\
      \xtopPulses  = \xsort(\xoccurences(g, \xstrata)) \\
      b = \xstartPositions(\#\xtopPulses) \compose \xtopPulses \\
    }{
      \xgetIndispensability(g)  = \xdescend (g, b)\rinv\compose (\# g - \_) \\
    }\\[4ex]

    \xfrac{
      f = \xmin g\limg \{\ran S \} \rimg \\
      T = \xoccurences(g, f) \setminus \ran S\\
      U = \xoccurences(g, f-1)\\
      V = \xif~\# T = 0 ~\xthen~U~\xelse~T\\
  X = \big(S \compose (\xcyclicSucc (V \xcup \ran S)\rinv) \big) \ranres V \\
    }{
      \xdescend(g, S) \\ = \xif~ V \not= \emptyset ~\xthen~
      \xdescend(g, S\chop \xsquash(X)) \\
      ~\xelse~S \\
      } \\
  \end{array}$ 
\caption{Härpfer's Extended Indispensability Algorithm, generalized to
  \xGNSM{} and \xMNSM.}
  \label{tab-gen-algo3}
\end{table}

\xumbi{\newpage}
\subsection{Generalized Indispensability Algorithm}

The formalization in Table~\ref{tab-algo3}
closely follows the operation of the C++ algorithm in \cite{haerpfer}.
During redactional work it turned out that a generalization
is much easier constructed with the opposite implementation: Not
to assign weights to metric positions, but to construct a sequence of
these positions, reflecting their priorities. The function
\xcombine{} is thus replaced by simple list concatenation $\_\chop\_$.
This is shown in Table~\ref{tab-gen-algo3}.

The major change is that initial values for arbitrary many 
top-level subtrees are required. This is done by the
function \xstartPositions{}, which delivers a sequential order
reflecting their priorities.
(Technically the compose operator $\compose$ is used to permutate
the sorted sequence of the positions with highest metric weight,
to get the starting sequence.)

The function \xdescend{} gets the \xGNSM{} or \xMNSM{} value
and the sequence $S$ as built so far.
Again $f$ is the lowest metric level already assigned  (contained
in the range of $S$). $T$ are the positions which carry the
same value and are not yet assigned. If there are no such
positions, $U$ is taken for further processing, which are
the positions with the next lower weight.

A new sequence $X$ is built by replacing all values in $S$ by their
cyclic predecessor among all assigned and to-be-assigned positions.
This sequence is thinned out to a relation targeting only
the non-assigned positions.
Thus \xsquash{} must be applied to get a sequence
which is simply appended to the input.

\xdescend{} finally delivers a permutation of the domain of $g$
(because $U$ contains positions which are not yet in $S$
and which are appended exactly once to $S$. A new $U$ is
calculated until it is empty.)
This permutation is translated into the indispensability map by
inversion of the relation and of the numeric values.
(The lowest index in the resulting sequence represents the highest
indispensability.)

A formal proof that the algorithm from Table~\ref{tab-gen-algo3}
implies the results from Table~\ref{tab-algo3} is still open.

The generalization is perfectly sensible in the formal sense, but
it is still unclear what the \emph{musical significance} of this
approach can be.

\bibliography{indisp000}

\appendix
\section{Mathematical Notation}
\label{txt_notation}
\small

The employed mathematical notation is fairly standard, inspired by the
Z notation \cite{spivey-z}. 
The following table lists some details:

\begin{tabular}{l@{\quad}p{0.7\columnwidth}}

$\nat$ & All natural numbers, incl.\ Zero.\\

$\{ a \ldots \}$ & with $a$ an integer number: All integers equal or
  greater than $a$ (our extension).\\

$\# A$ & The cardinality of a finite set = the natural number of the
  elements contained.\\

$\power{} A$ & Power set, 
the type of all subsets of the set $A$, incl.\ infinites.\\

$A \setminus B$ & The set containing all elements of $A$ which are not in
  $B$.\\

$A \times B$ & The product type of two sets $A$ and $B$, i.e.\ all pairs 
     $\{c=(a,b) | a \in A \land b\in B\}$. \\

$A \fun B$ & The type of the \emph{total} functions from $A$ to $B$.\\

$A \pfun B$ & The type of the \emph{partial} functions from $A$ to $B$.\\

$r\rinv$ & The inverse of a relation \\

$f~\limg~s~\rimg$ & The image of the set $s$ under the function or relation
  $f$\\

$r\compose s $& The composition of two relations: the smallest relation
  s.t.  $a \mathrel r b \land b \mathrel s c \Rightarrow a \mathrel{(r
    \compose s)} c$.  (first apply $r$, then apply $s$) \\

$r^n$ & Relation resulting from $n$ times applying $r$.\\

  $\dom A, \ran A$ & Domain and range of a relation.\\

  $S \domres R$&$= R \cap (S\times \ran R)$, i.e.\ domain restriction of
  a relation.\\

  $R \ranres S$&$= R \cap (\dom R\times S)$, i.e.\ range restriction of
  a relation.\\

  $r \oplus s$& Overriding of function or relation  $r$ by $s$. \\
 & Pairs from $r$ are shadowed by pairs from $s$: \\
 & $r \oplus s = \bigl(r \setminus (\dom s \times \ran r)\bigr) \xcup s$,
 with \dom{} and \ran{} being domain and range, resp.\\

$\xseq~A$ & The type of finite sequences from elements of $A$,
i.e.\ of maps $\nat\pfun A$ with a contiguous range $\{0\dotdot n\}$ as its
domain. (Our variant; standard Z toolkit sequences are 1-based.) \\

$\alpha\chop\beta$ &Concatenation of two lists.\\

$\xsquash(a)$ & Turns any partial function $\nat\pfun A$ into a
$\mathsf{seq}~A$ by compactifying the indices. \\
  & (To allow $\nat$ including 0 is our extension.) \\

$\xhead(a)$ & The first  element in a sequence  $= a(1)$. \\

$\xtail(a)$ & All elements except the first.\\

$\xmin~A$ & The minimum from a set of values.\\

$\xID_A$ & $=\{a\in A \bullet (a, a)\}$, the identity relation.\\

$(\_ + 1)$ & Abbreviated notation of the lambda expression
   $\lambda x \bullet x + 1$. Our extension.\\

\end{tabular}\\[3ex]

The frequently used notation
\[ \xfrac{a\\b\quad\quad c}{d}\]
means as usual  $a\land b\land c~\Longrightarrow~d$.
Nearly always it should be read as an \emph{algorithm}:
``For to calculate $d$, 
try to calculate $a$, $b$ and $c$. If this succeeds, the answer $d$ is valid.''

Functions are considered as special relations; relations as sets of pairs.
So with functions, expressions like ``$f\xcup g$'' are defined.

\end{document}